\documentclass[doublecol]{epl2}

\usepackage{array}
\usepackage{float}
\usepackage{amsmath}
\usepackage{graphicx}
\usepackage[position=t,singlelinecheck=off]{subfig}
\usepackage{caption}
\usepackage{color}

\title{Three-Dimensional Autonomous Pacemaker in the Photosensitive Belousov-Zhabotinsky medium}
\shorttitle{Autonomous Pacemaker} 

\author{A. Azhand, J. F. Totz, \and H. Engel}
\shortauthor{A. Azhand \etal}

\institute{
  Institut f\"ur Theoretische Physik Technische Universit\"at Berlin -  Hardenbergstrasse 36, D-10623 Berlin, Germany
}

\abstract{
In experiments with the photosensitive Belousov-Zhabotinsky reaction (PBZR) we found a stable three-dimensional
organizing center that periodically emits trigger waves of chemical concentration. The
experiments are performed in a parameter regime with negative line tension using an open gel reactor
to maintain stationary non-equilibrium conditions.
The observed periodic wave source is formed by a scroll ring stabilized due to its interaction with a
no-flux boundary. Sufficiently far from the boundary, the scroll ring expands and undergoes
the negative line tension instability before it finally develops into scroll wave turbulence.
Our experimental results are reproduced by numerical integration of the modified Oregonator 
model for the PBZR. Stationary and breathing self-organized pacemakers have been found 
in these numerical simulations. In the latter case, both the radius of the scroll ring and the 
distance of its filament plane to the no-flux boundary after some transient undergo undamped 
stable limit cycle oscillations. So far, in contrary to their stationary counterpart, the numerically 
predicted breathing autonomous pacemaker has not been observed in the chemical experiment.
}

\begin{document}

\maketitle

\section{Introduction}

Target patterns arising in response to periodic wave emission from a spatially localized organizing center are well-known examples for spontaneous formation of spatio-temporal structures in oscillatory or excitable reaction-diffusion (RD) systems. They were among the first patterns observed in the Belousov-Zhabotinsky (BZ) reaction \cite{Zaikin:Nature:1970}. Later, target patterns have been described in a large variety of RD systems of quite different chemical, physical and biological origin \cite{Jakubith:PRL:1990, Assenheimer:PRL:1993, Astrov:PRL:1998, Blasius:Nature:1999, Christoph:JChemPhys:1999}.

The wave source in the center of the target pattern that periodically emits trigger waves is called pacemaker (PM). Two types of PM have been observed. First, a PM can be connected with a local heterogeneity in the medium supporting wave propagation. Kuramoto has studied formation of target patterns in oscillatory media due to a localized frequency defect \cite{Kuramoto:Book:2003}. Target pattern result when uniform oscillations or phase waves within a localized oscillatory region propagate as trigger waves into a surrounding excitable medium. A well-known example for this kind of PM is the sinus node in the heart \cite{Jalife:CircRes:1986}.

\begin{figure*}[!t]
\begin{centering}
\includegraphics[width=1\linewidth]{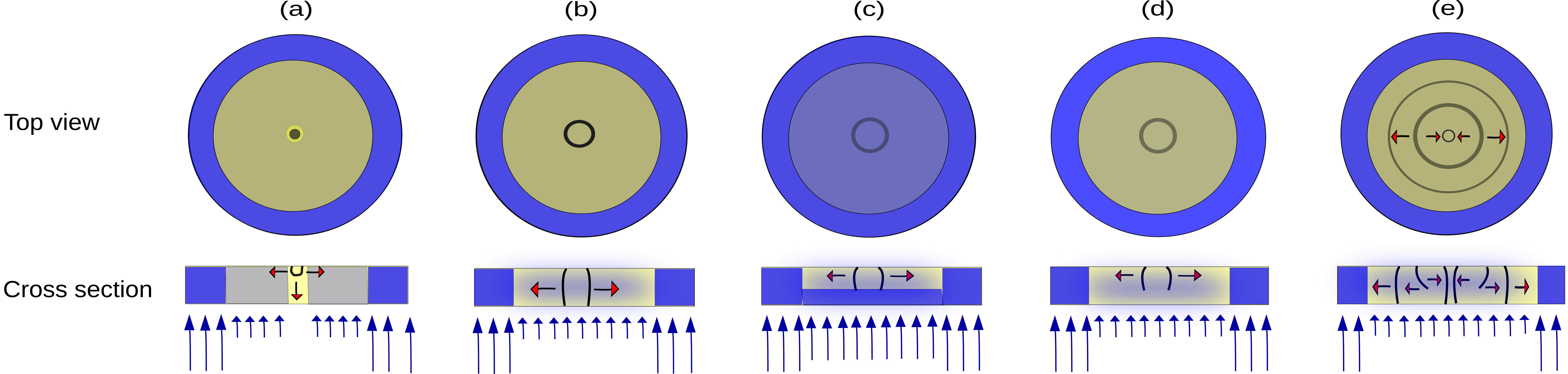}
\par\end{centering}

\caption{(Colour online) Optical initiation of a SR in a PBZ medium. The catalyst-loaded gel disk (diameter $5\, cm$) is illuminated as shown in the cross section by blue arrows whose length is proportional to the incident light intensity. Red arrows indicate the direction of wave propagation. Top view: dark blue -- unexcitable outer ring to prevent undesirable wave nucleation, light-blue -- partial wave annihilation, ocher -- excitable domain. The yellow region in the cross section (a) corresponds to the initially dark central region with spontaneously emerging phase waves (explanation see text).
\label{fig:ExpRingInit}}
\end{figure*}

Second, there exist PMs not pinned to a heterogeneity but resulting from intrinsic dynamical processes in spatially uniform RD media. These self-organized PM are called autonomous pacemakers (APM). Already in 1989, Vidal et al. and Nasumo et al. reported evidence for APM \cite{Vidal:JPhysChem:1989, Nasumo:JPhysSocJpn:1989}. APMs have been modeled in several oscillatory RD systems \cite{Mikhailov:PhysicaD:1992, Sakaguchi:ProgTheorPhys:1992, Rovinski:PhysRevE:1997}.
Mikhailov and Stich have shown analytically that while a stable APM cannot exist in oscillatory RD systems close to a supercritical Hopf bifurcation, they are possible in the vicinity of a codimension-2 pitchfork-Hopf bifurcation \cite{Stich:PRL:2001}. In the latter case, the reaction kinetics displays birhythmicity, i.e., coexistence of two stable limit cycles. Additionally, these authors presented a three-component activator-inhibitor system based on the FitzHugh-Nagumo model that gives rise to a stable APM in the excitable regime. Panfilov et al. found that mechanical deformation can induce an APM in a coupled reaction-diffusion-mechanics system \cite{Panfilov:PRL:2005}. Recently, Postnov et al. have reported APMs in a three-component phenomenological RD model of spreading depression \cite{Postnov:BrainResearch:2012}. Mutual annihilation of a pair of counterrotating spirals is another way to create a target pattern \cite{Mueller:PhysicaA:1992}.

By definition target patterns refer to two-dimensional (2d) systems. The majority of active media is three-dimensional (3d) and often spatially confined. Recently, 3d generalizations of rotating excitation waves in RD systems called scroll waves (SW) have attracted much interest. In this paper, we will focus on scroll rings (SR). SWs and SRs have been studied experimentally by computer tomography \cite{Winfree:Chaos:1996, Bansagi:PRL:2006, Bansagi:PRE:2007, Luengviriya:PRL:2008, Daehmlow:PRL:2013} as well as analyzed theoretically \cite{Yakushevich:1984, Keener:PhysicaD:1988, Biktashev:PhilTransRSocA:1994, Dierckx:PRE:2013} and by computer simulations \cite{Amemiya:Chaos:1998, Biktashev:IntJBifurcatChaos:1998, Alonso:Science:2003, Zaritski:PRL:2004, Alonso:JPhysChemA:2006, Alonso:PRL:2008, Jimenez:EPL:2010}.

The paper is organized as follows. In the next section we introduce briefly the chemical experiment and present the experimental results. Our experimental findings agree well with 3d numerical simulations carried out with the modified Orgeonator model for the PBZR, as will be discussed in section 3. We summarize our results in the concluding section 4.

\section{Experiment}

\begin{figure*}[!t]
\begin{centering}
\includegraphics[width=1\linewidth]{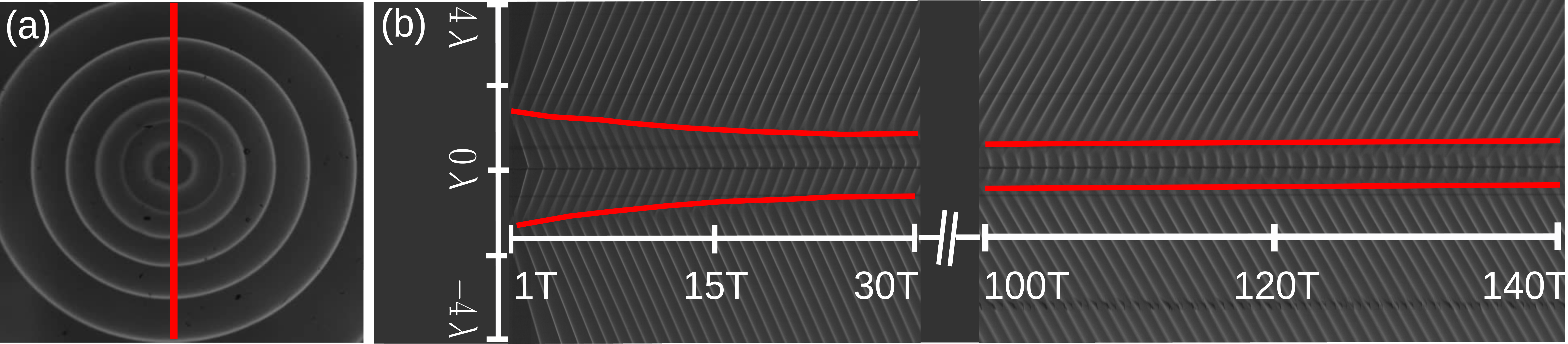}
\par\end{centering}

\caption{(Colour online) Stationary APM observed in the chemical experiment. (a) Snapshot obtained 15 periods of spiral wave rotation after SR initiation (top view). (b) Space–time plot along the red horizontal line in (a). Red lines in (b) indicate the location of the SR filament.
\label{fig:ExpPacemaker_Xt}}
\end{figure*}

Especially when the photosensitive catalyst is immobilized in a gel matrix to inhibit convection, and when the experiments are run in an open reactor to maintain stationary non-equilibrium constraints, the PBZR is very suitable to study the dynamics of RD waves under controlled laboratory conditions \cite{Braune:CPL:1993a, Krug:JPhysChem:1995, Kadar:JPhysChemA:97a, Brandtstaedter:CPL:2000}.

In our experiments we fixed the photosensitive catalytic complex Ruthenium-4,4'-dimethyl-2,2'-bipyridyl in a thin ($1.0-1.5\, mm$) layer of silica hydrogel with diameters of $5-6\, cm$. The concentration of the catalyst was $4.5\, mM$. Gel and catalyst preparation are described in detail in \cite{Brandtstaedter:CPL:2000}. The catalyst-loaded gel layer was placed vertically in the reactor chamber to allow $CO_2$ bubbles that form during the reaction to rise. The catalyst-free BZ mixture containing the concentrations $0.2M\, [NaBrO_3]$, $0.17M\, [\text{malonic acid}]$, $0,39M\, [H_{2}SO_4]$, and $0.09M\, [NaBr]$ was pumped continuously through the reactor at a rate of $50 ml/h$. The same composition has been used by Kheowan et al. in a study aimed at the measurement of the kinematical parameters of spiral waves in weakly excitable media \cite{Kheowan:PCCP:2001}. Circulating water from a thermostat keeps the temperature fixed at $22.0 \pm 0.5 ^{\circ}\text{C}$.

In the photosensitive variant of the BZ reaction, at a given composition of the BZ mixture the local kinetics depends on the intensity of illumination, $P$, applied to the reaction layer. Under low light intensity the reaction is oscillatory. Increasing $P$, at a certain threshold, $P_1$, a transition to excitable kinetics takes place. Further increase of $P$ raises the local excitation threshold until beyond a second intensity threshold, $P_2$, the medium ceases to support wave propagation.

In our experiments actinic light was always applied in the direction of the symmetry axis of the scroll ring, i.e., normal to the gel layer and the filament plane via a video projector (Casio XJ-A140V). The wave pattern is recorded from transmitted light by a CCD camera (ImagingSource Europe, Sony CCD chip). The recording light is spectrally separated from the actinic light by yellow band pass filters. Further details of the experimental set-up will be reported in a forthcoming paper.

One of the advantages of the PBZR is the elegant initiation of SWs exploiting wave inhibition in response to appropriate light application \cite{Panfilov:1984, Agladze:PhysicaD:1989, Linde:PhysicaD:1991, Amemiya:PRL77:1996}. This is illustrated for SR initiation in Fig. \ref{fig:ExpRingInit}. At the beginning, the projector provides the illumination pattern shown in panel (a). Between a strongly illuminated ($P > P_2$) outer ring and very low light intensity in a small circular central domain (yellow in Fig. \ref{fig:ExpRingInit},(a)), the gel layer is in the excitable regime ($P_1 < P < P_2$). The strongly illuminated outer ring inhibits parasitic wave nucleation at the periphery of the gel disc. After some time, a phase wave from the dark, oscillatory central region transforms into a cylindrical trigger wave propagating outwards (Fig. \ref{fig:ExpRingInit}, (b)). At a certain time moment we illuminate the gel layer (within the outer ring) uniformly with a light intensity $P>P_2$ that inhibits 
wave propagation and therefore eliminates the lower part of the cylindrical wave front as shown in Fig. \ref{fig:ExpRingInit}, (c). After restoring the intensity back to the level applied previously (Fig. \ref{fig:ExpRingInit}, (b)), the medium recovers excitability and the cut cylindrical wave curls up at its lower open end forming a circular closed filament (Fig. \ref{fig:ExpRingInit}, (d), (e)). Obviously, the radius, $R$, of the emerging SR and the distance of its filament plane from the layer boundaries, $z$, can be controlled varying the beginning of partial wave annihilation, and the intensity and/or the duration of illumination. Crucial for successful initiation of a planar filament plane oriented in parallel to the layer boundaries is, among other factors, an uniform illumination of the gel layer during partial annihilation of the cylindrical wave. Otherwise, scroll rings evolve with filament planes inclined with respect to the gel boundary or twist waves propagate along the filament. These cases will be not considered in the present paper.

\begin{figure}
\begin{centering}
\includegraphics[width=1\linewidth]{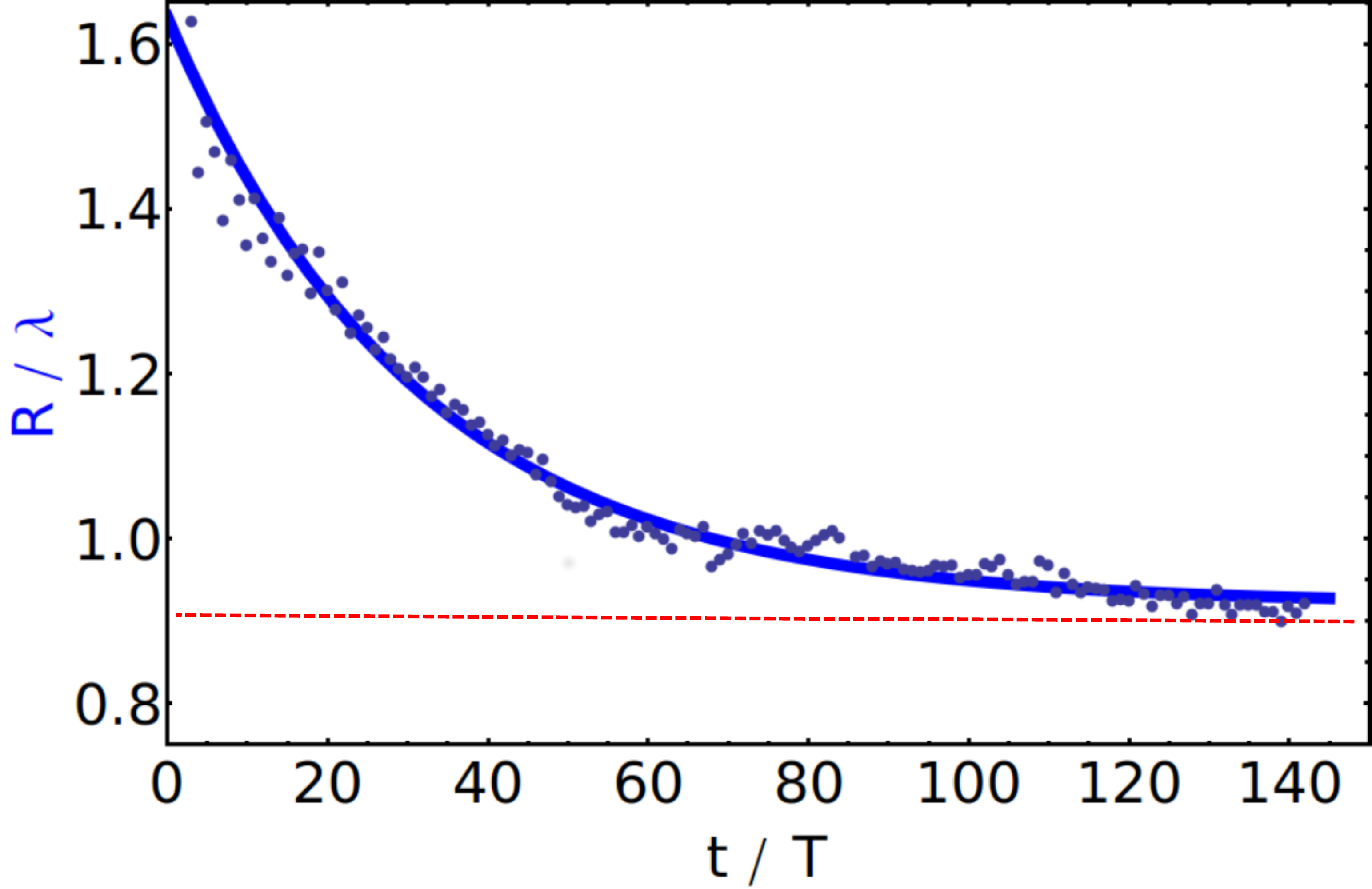}
\par\end{centering}

\caption{(Colour online) Radius of SR vs. time. The curve fits the points extracted from the experimental space-time plot in figure \ref{fig:ExpPacemaker_Xt} (b). From the very beginning, despite of negative filament tension the ring shrinks and asymptotically reaches a stationary radius (dashed red line). Thus, in the chemical experiment the SR forms a stable wave source acting as an APM over more than 140 rotation periods of the spiral waves forming the ring.
\label{fig:ExpPacemaker_Rt}}
\end{figure}

It is well-known that the level of illumination controls the rotation regime of spiral waves in 2d PBZR media. Usually, under increasing uniform illumination a cascade of transitions from rigid rotation via outward and inward meandering back to rigid rotation is observed (compare, for example, \cite{Brandtstaedter:CPL:2000}). Usually the transition from outward to inward meandering in 2d corresponds to a transition from positive to negative line tension in a 3d geometry \cite{Alonso:JPhysChemA:2006, Alonso:PRL:2008}. We have studied the evolution of SRs confined to the gel layer at constant applied illumination out of the interval $0.12\, mW/ cm^2 < P < 0.16\, mW/cm^2$ under otherwise fixed conditions (composition of the BZ mixture, width of the layer, etc). It turned out that crucial for SR dynamics is whether the SR is within interaction distance to one of the gel boundaries or not. For example, interaction with a Neumann boundary increases the life time of collapsing SRs in media with positive line tension significantly (compare \cite{Totz:arxiv:2014}). Even more surprising might be the boundary-mediated stabilization of expanding SRs in a medium with negative filament tension that will be discussed in the following. Note, that in an unbounded medium, the unconfined SRs would undergo a negative line tension instability finally leading to scroll wave turbulence \cite{Biktashev:IntJBifurcatChaos:1998, Alonso:Science:2003, Zaritski:PRL:2004}.

An example of a boundary-stabilized SR is shown in figures \ref{fig:ExpPacemaker_Xt} and \ref{fig:ExpPacemaker_Rt}. This experiment was performed at constant light intensity $P = 0.14\, mW/cm^2$. At the same intensity, Kheowan et al. have measured for rigidly rotating spiral waves a wavelength $\lambda \approx 2.3\, mm$, a core diameter $d \approx 0.5\, mm$, and a rotation period $T \approx 60\, s$  \cite{Kheowan:PCCP:2001}. These values correspond to very low excitability and negative filament tension (see for example \cite{Alonso:JPhysChemA:2006, Alonso:PRL:2008}). The overall thickness of the gel layer was approximately $1.4\, mm \approx 0.6\, \lambda$. As a consequence, the studied SR was located within interaction distance to the upper and lower layer boundary and therefore definitely confined. Because in our experiments the gel layers are placed on top of a glass plate, the lower boundary can be seen as a no-flux boundary. The diameter of the gel layer $~ 5\, cm \approx 22\, \lambda$ was large enough to exclude interaction of the centrally placed SR with the lateral layer boundaries.

\section{Numerical Simulations With the Modified Oregonator Model for the PBZR}

In 1990 Krug et al. proposed a modification of the Field-K{\"o}r{\"o}s-Noyes mechanism that accounts for the influence of oxygen and light on the BZ reaction \cite{Krug:JPhysChem:90}. This model has been used by many authors for the description of experimental results obtained in PBZ media (compare, for example, \cite{Braune:CPL:1993a, Krug:JPhysChem:1995, Kadar:JPhysChemA:97a}). In dimensionless form the modified Oregonator equations read

\begin{align}
\frac{\partial u}{\partial t} & =\frac{1}{\epsilon_{u}}\left[u-u^{2}+w(q-u)\right]+\Delta u\nonumber \\
\frac{\partial v}{\partial t} & =u-v\label{eq:MCO}\\
\frac{\partial w}{\partial t} & =\frac{1}{\epsilon_{w}}\left[\phi+fv-w(q+u)\right]+\delta \Delta w,\nonumber
\end{align}

\begin{figure}
\begin{centering}
\includegraphics[width=1\columnwidth]{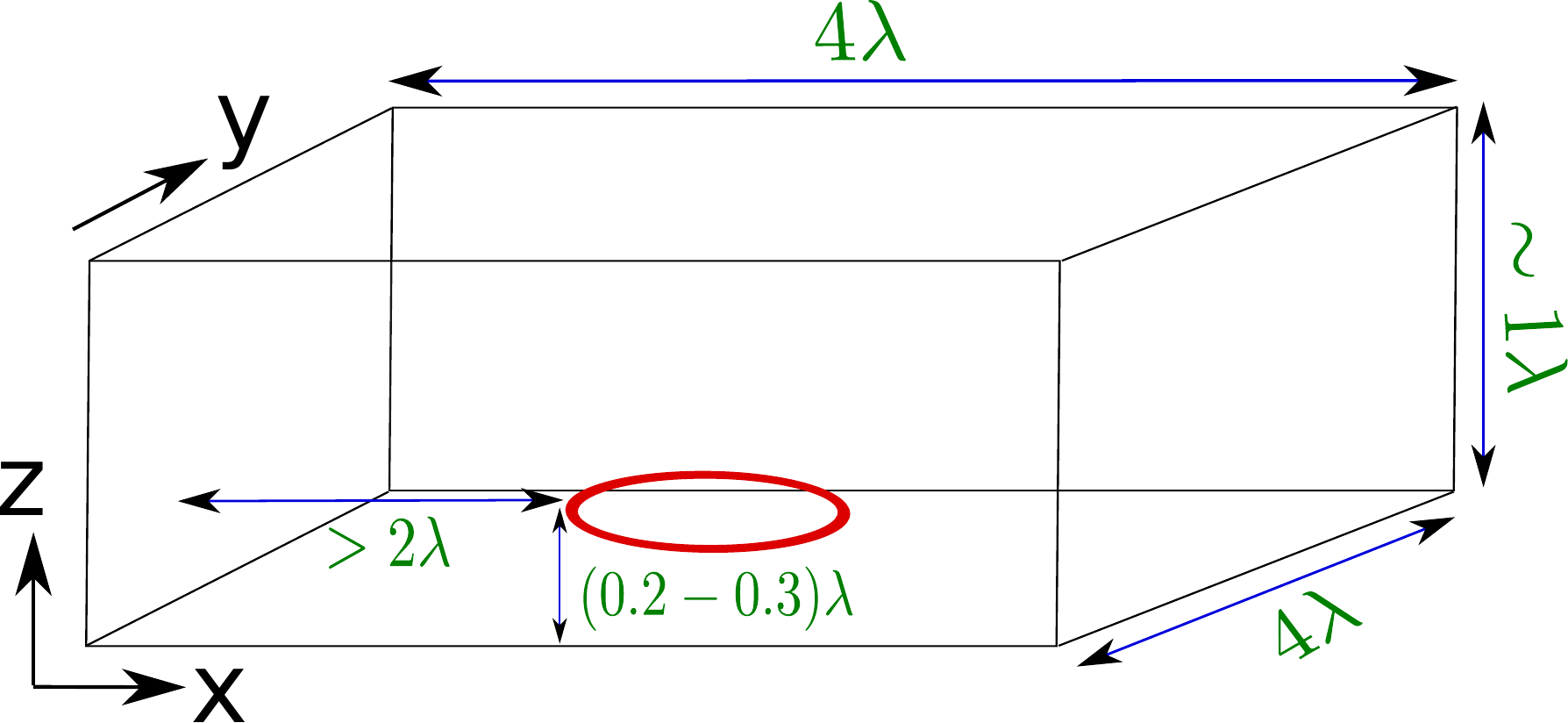}
\par\end{centering}

\caption{(Colour online) Rectangular box used for the numerical integration of the modified Oregonator equations with the SR filament in red. All boundaries are Neumann, for the shown SR placement, however, only the lower horizontal plane is within interaction range.
\label{fig:SimSpaceConditions}}
\end{figure}

Variables $u$, $v$, and $w$ are proportional to the concentrations $[HBrO_2]$ of bromous acid, $[Ru(bpy)^{3+}]$ of the oxidized catalyst (recorded experimentally), and $[Br^{-}]$ bromide, respectively. $\Delta$ denotes the 3d Laplacian. The ratio of the diffusion coefficients of activator $u$ (bromous acid) and inhibitor $w$ (bromide) was chosen to be $\delta =D_w/D_u = 1.12$.
There is no diffusion in the second equation as the catalyst is fixed in the gel matrix. Parameters $\epsilon _u$ and $\epsilon _w$ with $\epsilon _u \ll \epsilon _w$ describe the recipe-dependent time scales of the kinetics. In our simulations we will not eliminate the fast bromide kinetics adiabatically. This allows to account for bromide diffusion which is usually omitted or treated incorrectly in two-variable approximations of the complete model as given by Eqs. \ref{eq:MCO}. Parameter $q$ is a ratio of kinetic constants of the reactions described in the FKN mechanism, and $f$ is an adjustable stoichiometry parameter (compare \cite{Krug:JPhysChem:90} for details).

The bromide balance described by the third equation contains a bromide source, $\phi$, that is assumed to be proportional to the intensity of applied illumination $P$. In our numerical simulations, as in the chemical experiment, we will fix all parameters except of $\phi$. The following parameter values have been used: $\epsilon = 0.125$, $\epsilon ' = \epsilon /90$, $q = 0.002$, $f = 1.16$. For this parameter set with varying $\phi$-values the wave lengths, $\lambda$,  and the rotation periods, $T$, of 2d spiral waves were calculated. Below, all distances and time intervals will be expressed in units of $\lambda$ and $T$, respectively. For the chemical composition fixed in our calculations, the transition from oscillatory to excitable kinetics occurs at $\phi_1 = 0.006$; beyond $\phi_2 = 0.022$ the medium becomes unexcitable. These two $\phi$ thresholds correspond to the threshold intensities $P_1$ and $P_2$ introduced in the previous chapter. As in the chemical experiment we have focused on a parameter 
range with weakly excitable kinetics, rigidly rotating spiral waves, and negative line tension.

\begin{figure*}[!t]
  \captionsetup[subfigure]{labelformat=empty}
  \centering
  \subfloat[\textbf{(a)}]{\includegraphics[width=1\linewidth]{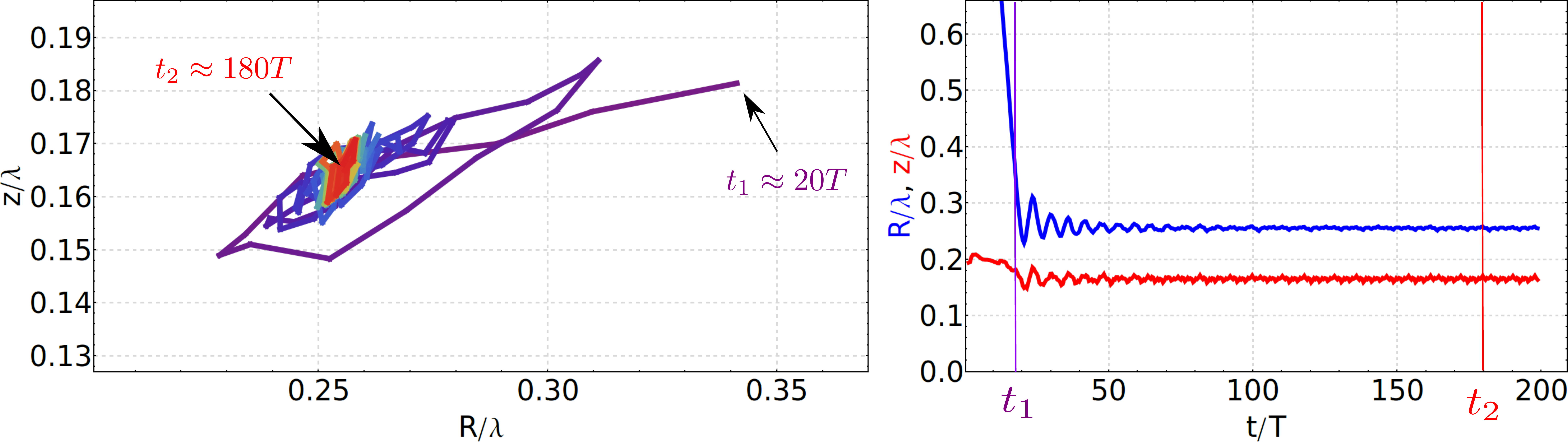}}\\
  \subfloat[\textbf{(b)}]{\includegraphics[width=1\linewidth]{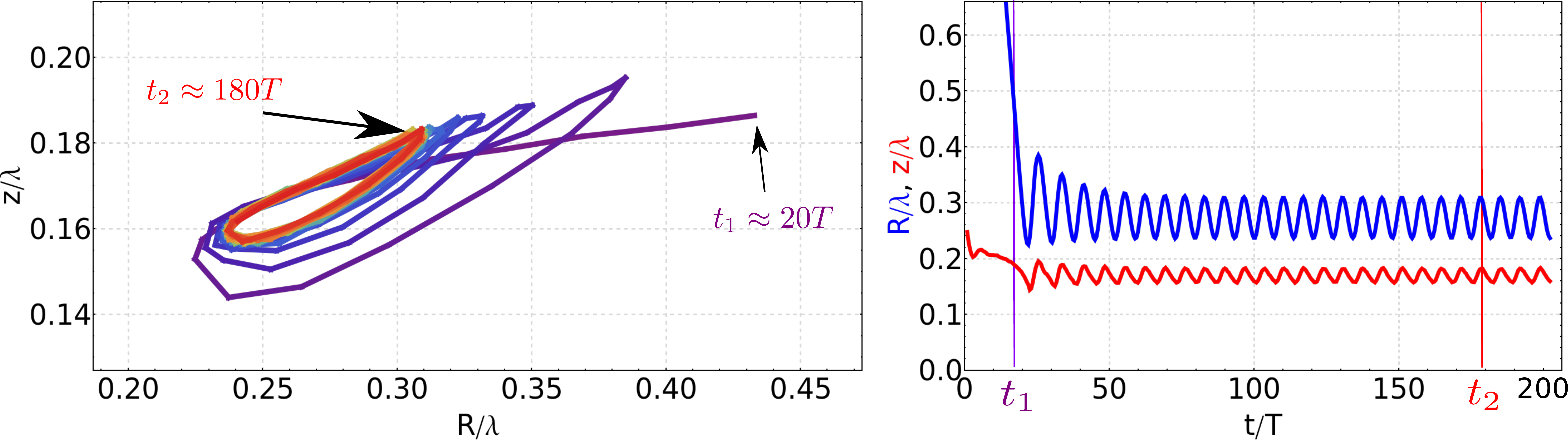}}
  \caption{(Colour online) Stationary (a) and breathing (b) APM obtained by numerical integration of the modified Oregonator equations for $\phi = 0.013$ and $\phi = 0.0125$, respectively. Left panels show the ring radius $R$ versus the $z$-coordinate of the filament plane, both quantities averaged over T. Color coding of the trajectory from $t_1$ (purple) to $t_2$ (red). Right panels plot $R$ (blue) and $z$ (red) as a function of time. Space and time are scaled in units of the wavelength and rotation period, respectively, of spiral waves forming the SR.
  \label{fig:DiffAutonomousPacemaker}}
\end{figure*}

We integrated Eqs. \ref{eq:MCO} in a rectangular cubic stripe (Fig. \ref{fig:SimSpaceConditions}) of size $x \times y \times z = 4 \, \lambda \times 4 \, \lambda \times 1 \, \lambda$ imposing Neumann boundary conditions on all lateral surfaces. A forward Euler scheme for time integration and a nineteen point star discretization for the Laplacian were applied with time and space discretization equal to $dt = 0.001$ and $dx = dy = dz = 0.3$, respectively.

In the numerical simulations the SR was initialized as in the chemical experiment. An outwardly propagating cylindrical wave front (cylinder axis in $z$-direction) was exposed to "inhibiting illumination" by increasing the photochemically induced bromide flow $\phi$ for a certain time interval $\Delta t_{\text{ini}}$ to a value $\phi _{\text{ini}} \gg \phi_2$ in the complete spatial region from $z=a$ to $z=z_{\text{top}}$. The filaments' initial distance $z_0$ from the bottom of the layer and the initial radius $R_0$ can be controlled choosing appropriate values of 
the parameters $\Delta t_{\text{ini}}$ and $a$. This procedure annihilates the wave in the "illuminated region" and leads to the formation of an untwisted SR with a filament plane exactly in parallel to the $x$-$y$ plane, after the medium has recovered to the $\phi$ level before inhibition (compare Fig. \ref{fig:SimSpaceConditions}). We emphasize that the position of the emerging scroll ring excludes interaction with the lateral Neumann boundaries which are at least $2 \, \lambda$ away from the filament.
The instantaneous position of the filament was defined from the crossing of the two iso-concentration surfaces $u_f  = 0.3$ and $v_f  = 0.1$, eliminating an oscillatory contribution due to rigid rotation by averaging over one period $T$ of spiral waves forming the ring.

Before we studied the effect of the interaction with the lower layer boundary on ring dynamics, we have checked in numerical simulations that for the chosen parameters SRs undergo a negative line tension instability in an unbounded ($x \times y \times z = 4\, \lambda \times 4\, \lambda \times 4\, \lambda$) medium. The results of the numerical integration of the modified Oregonator equations for a SR interacting with the Neumann boundary for the same parameters are summarized in Fig. \ref{fig:DiffAutonomousPacemaker}. We found a stationary APM in simulations with $\phi= 1,3\times 10^{-2}$, for example (see Fig. \ref{fig:DiffAutonomousPacemaker}a). Despite of negative line tension, a SR with initial radius $R(t=0) \approx 1.5\, \lambda$ and a filament plane  at an initial distance $z(t=0) \approx 0.2\, \lambda$ from the no-flux boundary contracts rapidly (Fig. \ref{fig:DiffAutonomousPacemaker}a, right), and, after some transient, reaches stationary values for $R$ and $z$. Thus, finally a stationary APM is formed as 
observed in the chemical experiment. Qualitatively the same results have been obtained for $\phi = 1.4 \times 10^{-2}$ (not shown here).

The stable fixed point in Fig. \ref{fig:DiffAutonomousPacemaker}a (left) can undergo a Hopf bifurcation giving rise to limit cycle oscillations. Actually we verified oscillatory behavior for $\phi = 1.2 \times 10^{-2}$ and  $\phi= 1.25 \times 10^{-2}$. The result for the latter case is shown on Fig. \ref{fig:DiffAutonomousPacemaker}b. Now, in the asymptotic regime both $R$ and $z$ display undamped oscillations: $R$ grows and the SR departs from the no-flux boundary. At some maximum $R$ value the ring starts to contract and simultaneously approaches the boundary again. This cycle out of growth under repulsion followed by contraction under attraction was observed stably over more than $400$ rotation periods of the spiral waves (for convenience, in Fig. 5 only the first $200$ rotation periods are shown). The breathing period, $T_B$, revealed to be larger than the period of spiral rotation. At $\phi _2 = 1.25 \times 10^{-2}$, we found $T_B \approx 7\, T$, for example. In numerical simulations, the parameter interval of $\phi$-values with stationary APM covers approximately $7/8$ of the whole existence range of APM. So far, the breathing APM was not found in the chemical experiment.

In all simulations both the stationary and the breathing pacemaker exhibited radii between $0.2$ and $0.3\, \lambda$ and were located at distances to the no-flux boundary varying between $0.15$ and $0.2\, \lambda$. Thus, the characteristic length scale for the interaction with the no-flux boundary is of the same order as the core size of the spiral wave \cite{Brandtstaedter:CPL:2000, Aranson:PhysRevE:1993, Aranson:PhysRevE:1994, HenryHakim:PhysRevE:2002, Biktasheva:PhysRevE:2003, Biktashev:PhysRevE:2009, Kupitz:JPhysChemA:2013}.

\section{Summary}

In this paper we report on the (to our knowledge first) experimental observation of a stable SR in an excitable medium with negative filament tension. The SR is stabilized by the interaction with a confining no-flux boundary which inhibits the negative line tension instability and the development of scroll wave turbulence that unavoidably would take place in an unbounded medium. Instead, a stable stationary APM emerges that represents the experimental verification of a stable three-dimensional organizing center predicted 25 years ago by Nandapurkar and Winfree in numerical simulations of spatially confined FitzHugh-Nagumo dynamics \cite{Nandapurkar:PhysicaD:1989}. Thus, spatial confinement can create stable APMs in excitable media.

The observed boundary-stabilized SR adds a new type of APM that is likely to exist in a large variety of excitable media of different physical, chemical, and biological origin. For example, in numerical simulations with the three-component Fenton-Karma model describing the propagation of action potential in cardiac tissue APM was found close to a Neumann boundary \cite{Fenton:Thesis:1999}. Thus, our results might be interesting to the understanding of self-organized pacemaker activity in comparatively thin layered parts of atrial myocardium.

Stabilization of the scroll ring can be roughly understood as follows. When the scroll ring interacts with a no-flux boundary, boundary-induced drift of the spiral waves forming the ring is super-imposed to the intrinsic expansion the scroll ring undergoes in an unbounded medium. Under certain conditions, both effects will compensate each other stabilizing the scroll ring at a finite radius. A more elaborated explanation of the stabilization mechanism based on a kinematical model that combines intrinsic and boundary-induced dynamics and predicts both the stationary as well as the breathing APM will be proposed in a separate paper.

\acknowledgments
We acknowledge financial support from the German Science Foundation DFG within SFB 910.

\end{document}